\documentclass[fleqn,usenatbib]{mnras}

\usepackage{newtxtext,newtxmath}
\usepackage[T1]{fontenc}
 
\usepackage{hyperref}
\usepackage{graphicx}	
\usepackage{amsmath}	
\usepackage{amssymb}	
\usepackage{gensymb}
\usepackage{caption}
\usepackage[dvipsnames]{xcolor}  

\newcommand{\reb}{{\sc \tt REBOUND}\:}

\newcommand{\whfast}{{\sc \tt WHFast}\:}

\title[Stability Constrained Characterization]{Stability Constrained Characterization of Multiplanet Systems}
\graphicspath{{./}{figures/}}

\author[Tamayo, Gilbertson, Foreman-Mackey]{
Daniel Tamayo$^{1}$\thanks{NHFP Sagan Fellow: dtamayo@astro.princeton.edu}\thanks{Both authors contributed equally to this manuscript}
, Christian Gilbertson$^{2,3,4}$\thanks{chrisgil@psu.edu}\footnotemark[2], Daniel Foreman-Mackey$^{5}$
\\
$^{1}$Department of Astrophysical Sciences, Princeton University, Princeton, NJ 08544, USA \\
$^{2}$Department of Astronomy and Astrophysics, 525 Davey Laboratory, The Pennsylvania State University, University Park, PA 16802, USA\\
$^{3}$Center for Exoplanets \& Habitable Worlds, University Park, PA 16802, USA\\
$^{4}$Institute for Computational and Data Sciences, The Pennsylvania State University, University Park, PA, 16802, USA\\
$^{5}$Center for Computational Astrophysics, Flatiron Institute, New York, NY 10010, USA
}

\date{Accepted XXX. Received YYY; in original form ZZZ}

\pubyear{2020}

\begin{document}
\label{firstpage}
\pagerange{\pageref{firstpage}--\pageref{lastpage}}
\maketitle

\begin{abstract}
Many discovered multiplanet systems are tightly packed.
This implies that wide parameter ranges in masses and orbital elements can be dynamically unstable and ruled out.
We present a case study of Kepler-23, a compact three-planet system where constraints from stability, transit timing variations (TTVs), and transit durations can be directly compared.
We find that in this tightly packed system, stability can place upper limits on the masses and orbital eccentricities of the bodies that are comparable to or tighter than current state of the art methods.
Specifically, stability places 68\% upper limits on the orbital eccentricities of 0.09, 0.04, and 0.05 for planets $b$, $c$ and $d$, respectively.
These constraints correspond to radial velocity signals $\lesssim 20$ cm/s, are significantly tighter to those from transit durations, and comparable to those from TTVs.
Stability also yields 68\% upper limits on the masses of planets $b$, $c$ and $d$ of 2.2, 16.1, and 5.8 $M_\oplus$, respectively, which were competitive with TTV constraints for the inner and outer planets.
Performing this stability constrained characterization is computationally expensive with N-body integrations.
We show that SPOCK, the Stability of Planetary Orbital Configurations Klassifier,
is able to faithfully approximate the N-body results over 4000 times faster.
We argue that such stability constrained characterization of compact systems is a challenging ``needle-in-a-haystack" problem (requiring removal of 2500 unstable configurations for every stable one for our adopted priors) and we offer several practical recommendations for such stability analyses.
\end{abstract}




\color{black}

\section{Introduction} \label{sec:intro}

The continually growing sample of discovered exoplanet systems provides a rich laboratory in which to test theories of planet formation.
Combining mass measurements from radial velocities with radii inferred from transit data yields mass-radius relationships \citep{Weiss13, Weiss14, Wolfgang16, ChenKip16, Neil18, Ning18, Neil20}.
This has provided first constraints on the interior structure of planets beyond our solar system \citep{Valencia06, Fortney07, Seager07,  Rogers11, Lopez13, Howe15, Zeng16}.
Orbital eccentricities provide complementary information on the dynamical histories of planetary systems.
Interactions with their natal protoplanetary disks tend to circularize exoplanet orbits \citep{Ward88}, so eccentricities measured at the present day constrain the degree of dynamical excitation across a given system's lifetime \citep{Rasio96, Weidenschilling96, Lin97, Marzari02, Adams03, Chatterjee08, Juric08, Simbulan17}.

Unfortunately, measurement of planetary masses and especially orbital eccentricities is challenging for the low-mass planets that dominate the exoplanet distribution, particularly around the faint stars searched by the {\em Kepler} mission \citep{Borucki11}.
For the minority of exoplanets near mean-motion resonances \citep{Fabrycky14}, it is possible to extract extremely precise masses and eccentricities even for low-mass planets \citep{Agol05, Holman05, Nesvorny08, Lithwick12, Wu13, Hadden14, Jontof15, Jontof16, Hadden17}.
Additionally, for eccentric giant planets \citep{Dawson12}, or when strong density constraints on the host star are available \citep[e.g.,][]{vanEylen15}, orbital eccentricities can be estimated from transit durations in individual systems.
However, only a small fraction of the sub-Neptunes that dominate the current multiplanet sample have observational constraints on their masses ($\approx 18\%$) or orbital eccentricities ($\approx 11\%$)\footnote{2692 confirmed sub-Neptunes from the NASA exoplanet archive, taken as planets with a radius < 4 Earth radii ($R_\oplus$), or a mass (or $M\sin{i}$) $<20$ Earth masses ($M_\oplus$).}.

In this paper we explore the independent constraints provided from orbital stability considerations.
Many of the known multi-planet exoplanet systems are in dynamically packed configurations \citep[e.g.,][]{Lissauer11, Fang12}.
This implies that many combinations of masses and orbital eccentricities would lead to rapid dynamical instabilities in those systems. 
Given the low likelihood of discovery just prior to such a violent orbital rearrangement, one can thus reject wide areas of parameter space to constrain physical and orbital parameters that may otherwise be inaccessible observationally. 
We refer to this process as stability constrained characterization.

Several authors have incorporated such stability constraints through direct N-body integrations for important exoplanet discoveries \citep[e.g.,][]{Steffen13, Tamayo15, Tamayo17, Quarles17, Wang18, Rosenthal19}.
However, this brute-force approach is orders of magnitude too computationally expensive to effectively sample the high-dimensional parameter spaces for timescales comparable to the Gyr ages of most known multiplanet systems.
To alleviate the computational burden, authors therefore typically restrict themselves to checking for ``immediate" instabilities within $10^4-10^6$ orbits.
However, this is not ideal given that dynamical instabilities in typical super-Earth compact planetary systems tend to be approximately logarithmically spaced in time, with similar numbers of instabilities occurring in each decade of time out to at least billions of orbital timescales \citep[e.g.,][]{Volk15}.

There have been extensive efforts to predict such instabilities analytically
, yielding many powerful results for two-planet systems \citep{Wisdom80, Marchal82, Gladman93, Barnes06, Deck13, Petit17, Petit18, Hadden18, Hadden19}.
In the 3+ planet case, several authors have run N-body integrations with initial conditions drawn from low-dimensional cuts through the full parameter space, and fitted empirical functional forms to the resulting instability times \citep{Chambers96, Yoshinaga99, Marzari02, Zhou07, Smith09, Funk10, Pu15, Obertas17, Gratia19}. 
Such empirical fits provide insight into the dependencies on physical parameters, and complementary analytic investigations have clarified 
several aspects of the underlying dynamics \citep{Zhou07, Quillen11, Laskar17, Yalinewich19, Petit20}.
However, none of these models are yet able to make reliable stability predictions in general, compact 3+ planet systems, particularly ones near mean-motion resonances \citep{Tamayo20}.

Recently, \cite{Tamayo20} presented the Stability of Planetary Orbital Configurations Klassifier (SPOCK), a machine learning model capable of making reliable stability predictions over $10^9$ orbits across a wide variety of compact orbital configurations similar to those discovered by the {\em Kepler} and TESS missions, up to $10^5$ times faster than direct N-body integration. 
They argue that for such $\lesssim$ Neptune-mass planets, short-timescale ($<10^9$ orbit) instabilities are still dominantly driven specifically by the overlap of two-body mean motion resonances, as in the two-planet case \citep{Wisdom80, Obertas17}.
For a given set of initial conditions, they run a short ($10^4$ orbit) N-body integration (see also \citealt{Tamayo16}) to generate a set of ten dynamically motivated summary features, two of which involve the MEGNO chaos indicator \citep{Cincotta03}, and the remaining eight derived from analytical two-planet mean-motion resonance models \citep{Hadden19}.
These features are then passed to a gradient-boosted decision tree machine learning classifier \citep{Chen16}, which returns an estimated probability of stability over $10^9$ orbits.
\cite{Tamayo20} show that their model not only performs well on a holdout set of resonant training examples (not used during the training process), but also to non-resonant and higher-multiplicity systems.

By enabling long-term stability classification of such compact orbital configurations in a fraction of a second (compared to several hours for a $10^9$ orbit N-body integration), SPOCK computationally opens up the stability constrained characterization of multi-planet systems.
In this paper we demonstrate how one can use SPOCK to sharpen poorly constrained physical and orbital parameters of exoplanets and explore the particular combinations of physical and orbital parameters that stability specifically informs.

The paper is organized as follows.
We begin in Sec.\:\ref{preliminaries} with various definitions and a discussion of the implicit assumptions and possible biases introduced by stability constrained characterization.
In Sec.\:\ref{sec:nbody} we present stability constraints through direct N-body integrations, and contrast them against those from transit timing variations and transit durations in the compact 3-planet Kepler-23 system, where all methods can be directly compared.
In Sec.\:\ref{sec:nbodycomp} we consider faster stability constraints using SPOCK, provide practical recommendations for its use, and compare these estimates to the N-body results.
We conclude in Sec.\:\ref{conclusion}.

\section{Stability Constrained Characterization} \label{sec:preliminaries}

\subsection{Preliminaries and Definitions} \label{preliminaries}
Before applying stability constraints to planetary systems, we begin with some definitions and by laying out the underlying assumptions.

The orbital evolution of typical multi-planet systems (in particular of any configurations that would go unstable after a few dynamical cycles) is chaotic.
This leads to effectively stochastic evolution, most notably in the orbital eccentricities, which in simple models diffuse until orbits begin crossing and cause close encounters between the planets \citep[e.g.,][]{Murray97, Zhou07}.

Once orbits cross, it can take a long time for close-in low-mass planets to find one another and physically collide \citep{Rice18}.
However, we assume that in such crossing configurations, the large eccentricities and strong scatterings would be detectable either through transit photometry \citep[e.g.,][]{Ford08, Kipping12, Dawson12, vanEylen15, Price15} or by sharp ``chopping'' variations in their transit times due to close approaches \citep{Nesvorny14, Deck15}.  
Following previous authors \citep[e.g.,][]{Gladman93, Zhou07, Faber07, Smith09, Obertas17}, we therefore operationally define the ``instability time'' as the time it takes for a pair of planets to come within one Hill sphere of one another.
Once this occurs, orbits start crossing almost immediately (on orbital timescales, \citealt{Gladman93}), so this criterion is both accurate and simple to compute numerically with fast N-body algorithms \citep[e.g.,][]{Wisdom91}.

Instability times are most usefully expressed as a number of orbits.
Because point-source Newtonian gravity is scale invariant, we can put different systems on equal footing by expressing all masses relative to that of the central star, and all times and distances in units of the innermost planet's orbital period and semimajor axis, respectively. 
This facilitates comparisons between systems with different orbital periods and absolute ages.
For example, the $\sim 40$ Myr age of the HR 8799 system \citep{Marois08} is only one hundred times younger than the typical Gyr ages of observed systems.
However, the much longer orbital periods of these directly imaged planets $\sim 100$ yrs, as compared to $\sim 0.01-0.1$ yrs for typical exoplanets discovered through transits or radial velocities, means that the HR 8799 system is dynamically significantly younger than most known planetary systems. For example, HR 8799 is roughly only two hundred times younger than the $\sim 8$ Gyr old TRAPPIST-1 \citep{Burgasser17}, but the innermost HR 8799e has executed $\sim 10^6$ fewer orbits than the innermost TRAPPIST-1b, whose orbital period is only 1.5 days \citep{Gillon17}. 

The fact that the required timestep for N-body integrations scales linearly with the innermost orbital period puts suites of direct integrations over the $\sim 10^6$-orbit dynamical age of the HR 8799 system within computation reach \citep{Wang18}.
However, statistical exploration of the long-term stability of configurations of typical multi-transiting systems with dynamical ages of $\sim 10^{10}-10^{12}$ orbits becomes computationally prohibitive with N-body methods (a few CPU hours per billion orbits per configuration with the fastest available algorithm of \citealt{Wisdom91}).

\subsection{Implications of Chaotic Dynamics} \label{sec:chaos}
The chaotic dynamics leading to such instabilities have at least two important implications.
First, chaos renders Newton's time-reversible equations of motion effectively irreversible. 
Planets on crossing orbits will not scatter back onto circular orbits.
In particular, integrating a system's current orbital configuration backward in time does not reveal its past history beyond a few chaotic (Lyapunov) timescales.
Rounding errors due to finite floating point precision \citep[though see][]{Rein18}
renders integrations forward or backward in time statistically identical, with eccentricities diffusing upward in both time directions \citep{Gaspard05, Morbidelli20}.
Given this loss of past information, we focus on integrations forward in time.

Second, this chaos implies that small changes to the initial conditions will yield a range of equally valid instability times.
\cite{Rice18} and \cite{Hussain19} studied the distributions of such instability times in compact planetary configurations, finding them approximately lognormally distributed.
While different orbital configurations in the dataset analyzed by \cite{Hussain19} had mean instability times spanning four orders of magnitude ($\approx 10^4-10^8$ orbits), the {\it widths} of the lognormal distributions imprinted by the chaotic dynamics was much narrower, approximately 0.4 dex.
This can be understood as a consequence of chaotic random walks in action space \citep{Petit20}. 
A given orbital configuration thus has a well-defined mean instability time, and an N-body integration provides a single draw from a relatively narrow distribution around that value.
These widths quantify the errors on instability times reported from N-body integrations, set the fundamental limit on instability time predictions \citep{Tamayo20}, and provide important constraints on the dynamics leading to instability \citep{Hussain19, Petit20}.

\subsection{Implicit Assumptions on the Formation and Evolution of Planetary Systems}

Eliminating unstable orbital configurations implicitly makes assumptions about the formation and evolution of planetary systems.
On one extreme, suppose that the orbital architectures of planetary systems are effectively set during or shortly after the protoplanetary disk phase, and from then on remain stable for the age of the universe.
Under this assumption, one could straightforwardly rule out orbital configurations with orbital lifetimes shorter than a Hubble time.

By contrast, planetary systems may instead be continually destabilizing and rearranging themselves into progressively longer-lived configurations \citep[e.g.,][]{Laskar90, Volk15, Pu15, Izidoro17, Izidoro19}.
In the traditional paradigm of core accretion, the final stage of mass growth for sub-Neptunes occurs in a phase of giant impacts \citep{Goldreich04review, Hansen12, Hansen13, Dawson16, Macdonald20}, which could continue occurring over timescales comparable with the age of the system \citep[e.g.,][]{Volk15, Pu15}.
More detailed models incorporating the effects of pebble accretion and migration \citep[e.g.,][]{Bitsch19, Lambrechts19} tend to capture planets into chains of mean motion resonances.
In order to reconcile this with the observed paucity of systems in such MMRs, such resonant chains must destabilize over time \citep{Izidoro17, Izidoro19}, and again the tail of such collisions could continue to the present day.
Thus, both of these hypotheses would predict that systems should exist with short remaining lifetimes (i.e., compared to the system's age).
To our knowledge no such systems have been identified to date, though quantitative estimates of how many such unstable systems one should find are not yet clear.
Nevertheless, if such unstable systems might exist, one should clearly not rule out unstable configurations out of hand.

We are pursuing separately the quantitative question of how frequently one should expect to discover systems with a given remaining lifetime, but we can make a useful simplification.
One can always reliably rule out configurations in the limit of instability times much shorter than the age of the system. 
For example, one could confidently rule out configurations with lifetimes shorter than $10^6$ orbits in a system with an age of $10^{12}$ orbits.
One would either have to be extremely lucky to catch a system immediately prior to such a cataclysm, or such rearrangements would have to be constantly reoccurring. 
\cite{Hadden17} uniformly analyzed 55 multi-planet systems where TTVs allow characterization of the full orbital architecture, and find that the majority of their solutions for each system are stable over at least $10^6$ orbits\footnote{The exception being resonant chains \citep[see also][]{Mills16, Gillon17}, where a full exploration of the phase space with Markov Chain Monte Carlo becomes difficult \citep{Tamayo17}}.
The fact that no systems with short instability times are currently known strongly disfavors a scenario where planetary systems are continually rearranging on short timescales at the present day, and this qualitative picture agrees with N-body simulations of giant impact accretion \citep[e.g.,][]{Dawson16}.

In this work, we choose to uniformly discard orbital configurations with lifetimes shorter than $10^9$ orbits, which corresponds to fractions of a few times $10^{-4}$ to a few times $10^{-2}$ in typical systems with lifetimes of a few Gyr and innermost orbital periods of $0.01-0.1$ yrs.
This should be a useful rule-of-thumb for typical exoplanet systems with short period planets, but would for example not be applicable to the young HR 8799 planets, with a dynamical lifetime of only $10^6$ orbits \citep{Wang18}.

For simplicity, for the remainder of the paper we therefore refer to systems surviving for $10^9$ orbits as (long-term) {\it stable}, and ones that suffer close encounters in that timeframe as {\it unstable}.

\subsection{Stability Constrained Characterization} \label{sec:scc}

Most current efforts to estimate orbital parameters and masses from observational data employ Markov Chain Monte Carlo (MCMC) methods that are efficient in the higher dimensional parameter spaces spanned by multiplanet systems, and make it simple to evaluate confidence intervals and correlations \citep[e.g.,][]{Foreman13}.
One approach to stability constrained characterization would be to compute the stability probability at each step in an MCMC analysis.
Here, we instead apply the stability constraint as a post-processing step performed after generating a set of samples that do not incorporate stability.
This has the benefit that this procedure is not tightly coupled to a specific analysis pipeline and can be used with existing tool chains and workflows that exist for characterizing exoplanet systems.

The goal of any Bayesian characterization is to compute posterior weighted expectation integrals such as
\begin{equation}
\label{eq:expect}
E_{p(\theta\,|\,\mathrm{data})}[f(\theta)] = \int f(\theta)\,p(\theta\,|\,\mathrm{data})\,\mathrm{d}\theta
\end{equation}
where $\theta$ represents the set of planet parameters (mass, orbital period, eccentricity, etc.) and $f(\theta)$ is the target of our inference (see \citealt{Hogg18}, for example, for a more complete discussion).
Typically, the ``data'' in Equation~\ref{eq:expect} refers to an observational dataset such as a light curve or radial velocity curve.
But, here we aim to include an extra piece of data: we have observed this planetary system, suggesting that it is in a stable configuration.
We refer to the collected observational data as $X$ and the ``observation'' of stability as $q$.
With this notation, the relevant posterior probability is
\begin{equation}
\label{eq:posterior}
p(\theta\,|\,X,\,q) = \frac{p(\theta)\,p(X\,|\,\theta)\,p(q\,|\,\theta)}{p(X)\,p(q)}
\end{equation}
where, on the right hand side, we have made the reasonable assumption that the observed data and stability are independent conditioned on the physical parameters of the system.
In other words, the calculation of a light curve model (for example) depends only on the parameters of the system and not on whether or not those parameters are stable.

Substituting Equation~\ref{eq:posterior} into Equation~\ref{eq:expect}, we find
\begin{eqnarray}
E_{p(\theta\,|\,X,\,q)}[f(\theta)] &=& \int \left[ f(\theta)\,\frac{p(q\,|\,\theta)}{p(q)}\right]\,\frac{p(\theta)\,p(X\,|\,\theta)}{p(X)}\,\mathrm{d}\theta \nonumber\\
&=&\int \left[ f(\theta)\,\frac{p(q\,|\,\theta)}{p(q)}\right]\,p(\theta\,|\,X)\,\mathrm{d}\theta \quad.\label{eq:fullpost}
\end{eqnarray}
Now, if we have somehow (using MCMC or otherwise) generated samples $\theta^{(n)} \sim p(\theta\,|\,X)$ from the posterior probability density only conditioned on the observational data, we can approximate the integral in Equation~\ref{eq:fullpost} using the usual sampling approximation
\begin{equation}
\label{eq:sampapprox}
E_{p(\theta\,|\,X,\,q)}[f(\theta)] \approx \frac{\sum_{\theta^{(n)}} w^{(n)}\,f(\theta^{(n)})}{\sum_{\theta^{(n)}} w^{(n)}}
\end{equation}
where $w^{(n)} = p(q\,|\,\theta^{(n)})$, i.e., the probability the given sample is stable, and the factors of $p(q)$ have canceled.
In other words, the stability constraint can be incorporated by re-weighting the samples from an existing MCMC analysis using the stability probability calculated for each sample in the chain.

When performing stability constrained characterization through N-body integrations, 
$p(q\,|\,\theta)$ becomes a simple binary probability: either stable or unstable with probability one for a given set of parameters\footnote{In reality, the uncertainty arising from chaotic dynamics (Sec.\:\ref{sec:chaos}) implies that $p(q\,|\,\theta)$ is not a step function but has finite width. Nevertheless, because this width is small, as measured numerically by \cite{Hussain19}, a step function is a good approximation.}, reducing to a rejection of unstable configurations since
\begin{equation} \label{binaryclass}
p(q\,|\,\theta) = \left\{\begin{array}{ll}
1 & \mathrm{if\,\theta\,is\,a\,stable\,configuration} \\
0 & \mathrm{otherwise.}
\end{array}\right.
\end{equation}
By contrast, we advocate for using the continuous probability of stability $p(q\,|\,\theta)$ estimated by SPOCK (Sec.\:\ref{sec:probabilistic}).

Equation~\ref{eq:sampapprox} becomes exact in the limit of infinite samples.
However, rejecting unstable configurations has the potential to substantially decrease the effective number of samples, and hence increase the error introduced by the sampling approximation.
This could happen, for example, if the stability constraint is significantly more constraining than the observational data.
Therefore, when using this method, care should be taken to ensure that the effective number of samples is sufficient to support the results.
In this paper, the initial samples $\theta^{(n)}$ are independent by design so it is sufficient to just make sure that the rejection step leaves a sufficiently large number of samples, but a more thorough analysis would be required when accounting for autocorrelation in MCMC analyses.

\section{N-body Stability Constraints} \label{sec:nbody}

We now consider the constraints imposed by requiring orbital stability, and compare the improvements to complementary observational methods. 
We focus on constraints from transit photometry that are immediately available in large transit surveys like Kepler and TESS, and later comment on radial velocity observations.
In particular, TTVs can provide information on the planetary masses \citep{Agol05, Holman05, Nesvorny08, Lithwick12, Wu13, Hadden14, Jontof15, Jontof16, Hadden17}, and both TTVs and transit duration modeling can constrain orbital eccentricities \citep{Dawson12, Kipping14, vanEylen15, Xie16}.

We choose the compact, near-resonant, and asteroseismicaclly characterized three-planet Kepler-23 system as a particularly illustrative example where stability, transit duration \citep{vanEylen15}, and TTV \citep{Hadden17} analyses can all be applied and compared. 

Kepler-23 (KOI 168) is a $1.078\pm0.077 M_\odot$, 13.4 Kepler-magnitude star \citep{Huber13}, with an estimated age of 4-8 Gyr \citep{Ford12}.
Its three known planets are tightly packed, with orbital periods of 7.107, 10.742 and 15.274 days (with periods ratios between adjacent planets of 1.511 and 1.422, respectively), and radii of $1.69 \pm 0.08, 3.12 \pm 0.10$ and $2.24 \pm 0.09$ Earth-radii ($R_\oplus$, respectively \citep{vanEylen15}.
Its dynamical age thus corresponds to $2-4 \times 10^{11}$ inner-planet orbits.

Precise stellar parameters from asteroseismology allow for eccentricity constraints from transit durations. A shorter-than-expected transit duration can be explained by its occurrence near the pericenter of an eccentric orbit. However, there is a degeneracy with the orientation of the pericenter relative to the line of sight (one can observe the same transit duration with a more eccentric orbit by adjusting the orbit orientation such that the transit occurs further from pericenter, see e.g., Fig. 1 of \citealt{vanEylen15}). Additionally, there is a degeneracy with the transit impact parameter, which will also change the transit duration, though this correlation can sometimes be alleviated by modeling the distortion of the transit shape with increasing impact parameter. These effects generically result in eccentricity constraints from transit durations having a tail toward large values \citep{vanEylen15}.

The period ratios between adjacent planets (1.511 and 1.422, respectively) fall near strong MMRs (3:2 and 7:5, respectively). 
This induces TTVs that have been measured and modeled to put constraints on the masses and orbital eccentricities \citep{Ford12, Hadden17}.
For these TTV and the above transit duration comparisons, we take the publicly available posterior distributions from \cite{Hadden17} and \cite{vanEylen15}, respectively.

Finally, we consider the constraints imposed by long-term stability.
As argued in Sec.\:\ref{sec:scc}, stability constraints are complementary and easily combined with observational constraints. 
However, for a straightforward comparison, we choose to present stability constraints independently.
Thus, rather than rejecting unstable posterior samples $\theta^{(n)} \sim p(\theta\,|\,X)$ as constrained by TTVs or transit durations (Sec\:\ref{sec:scc}), we reject unstable configurations starting from the same prior distribution of orbital configurations as considered by \cite{Hadden17} and \cite{vanEylen15} in their TTV and transit duration analyses.

The most reliable (but most computationally expensive) way of estimating stability is through direct N-body integration. 
We begin by comparing these ``ground truth'' results to the other methods, and in Sec.\:\ref{sec:nbodycomp} compare these N-body stability constraints to faster results obtained with SPOCK.

It is important to point out that constraints from each of the methods we compare  depend on a variety of physical and observational factors.
TTV constraints, in particular, can in some cases be much more precise than stability.
For reference, out of the 90 planet pairs uniformly analyzed by \cite{Hadden17}, Kepler-23 is roughly a typical case, with about 60\% of planet pairs yielding better eccentricity constraints (taken as the ratio of the reported error bars to the peak of the posterior probabilities). 

Transit durations are less informative for Kepler-23 than average among the sample selected by \cite{vanEylen15} with asteroseismology.
Approximately 70\% of their sample of $\approx 70$ planets have tighter 68th-percentile upper limits for the orbital eccentricity than those for Kepler-23.

By contrast, Kepler-23 is one of the more compact multiplanet systems known, which yields stronger eccentricity constraints from stability.
While it is difficult to quantify this precisely without an analytic understanding of such instabilities, only approximately 14\% of currently known multi-planet systems have a trio of planets with adjacent period ratios at or closer than the 3:2 MMR like Kepler-23. 
Nevertheless, Kepler-23 offers a rare and valuable opportunity to compare constraints from all three methods.

\subsection{Priors}

To facilitate comparisons, we follow \cite{Hadden17} and \cite{vanEylen15} by drawing all planet parameters independently, sampling eccentricities uniformly and masses log-uniformly. 
Eccentricities were drawn from [0,0.9], and masses by sampling bulk densities from 0.3-30 g/cc and using fixed planetary radii of 1.8, 3.2 and 2.3 $R_\oplus$ for planets, $b$, $c$ and $d$, respectively.
As typically the case for transiting planets, the orbital periods are measured precisely enough that for simplicity we fix them to the values listed above.
Particularly close to MMRs, these period ratios can matter for stability, so sampling them from a wider range would be prudent when periods are less certain, for example if there are transit timing variations over a small set of transits.

We sampled the longitudes of ascending node, arguments of pericenter and mean anomalies uniformly from $[0,2\pi]$, and inclinations uniformly from zero to values corresponding to an impact parameter of 0.9 stellar radii.
More carefully accounting for correlations between orbital eccentricities and impact parameters is important for some applications, e.g., modeling transit durations \citep{vanEylen15}.
By contrast, stability constraints are very weakly sensitive to small mutual inclinations.
Experimentation by \cite{Tamayo20} with inclination-dependent features provided negligible improvements to their machine learning models' performance on compact systems with mutual inclinations $\lesssim 10^\circ$, and their final SPOCK model has no features that depend on inclinations explicitly. 
Choices related to sampling inclinations should therefore be guided by the original orbital parameter inference problem, and should be negligible for stability constraints in near-coplanar systems over timescales of $10^9$ orbits.

Finally, we drew normally distributed stellar masses of $M_\star = 1\pm0.1 M_\odot$.
We note that as mentioned above, gravity's scale invariance implies that the dynamics only depend on the planet-star mass ratios, rather than their individual masses.
An alternative, therefore, that decouples uncertainties in stellar parameters would be to constrain planet-star mass ratios rather than absolute planetary masses.
The star's absolute mass does change the orbital periods $\propto M\star^{1/2}$, so one can convert from $10^9$ orbits to absolute time and from planet-star mass ratios to planetary masses by folding in the uncertainties in stellar mass.

We ran each configuration for $10^9$ orbital periods of the inner-most planet using a fixed timestep of 0.25 days $\approx$ 3.5\% of the innermost orbital period.
We used the integrator \whfast \citep{ReinTamayo15}, which is part of the open-source package \reb \citep{Rein12}, and is based on \cite{Wisdom91}. 
We reject any configurations where any two planets come within the sum of their individual Hill radii (see Sec.\:\ref{preliminaries}) within the span of integration, and accept samples that survive the full $10^9$ orbits.

\subsection{Constraints on Masses} \label{sec:mass}

\begin{figure*}
    \centering
    \resizebox{0.99\textwidth}{!}{\includegraphics{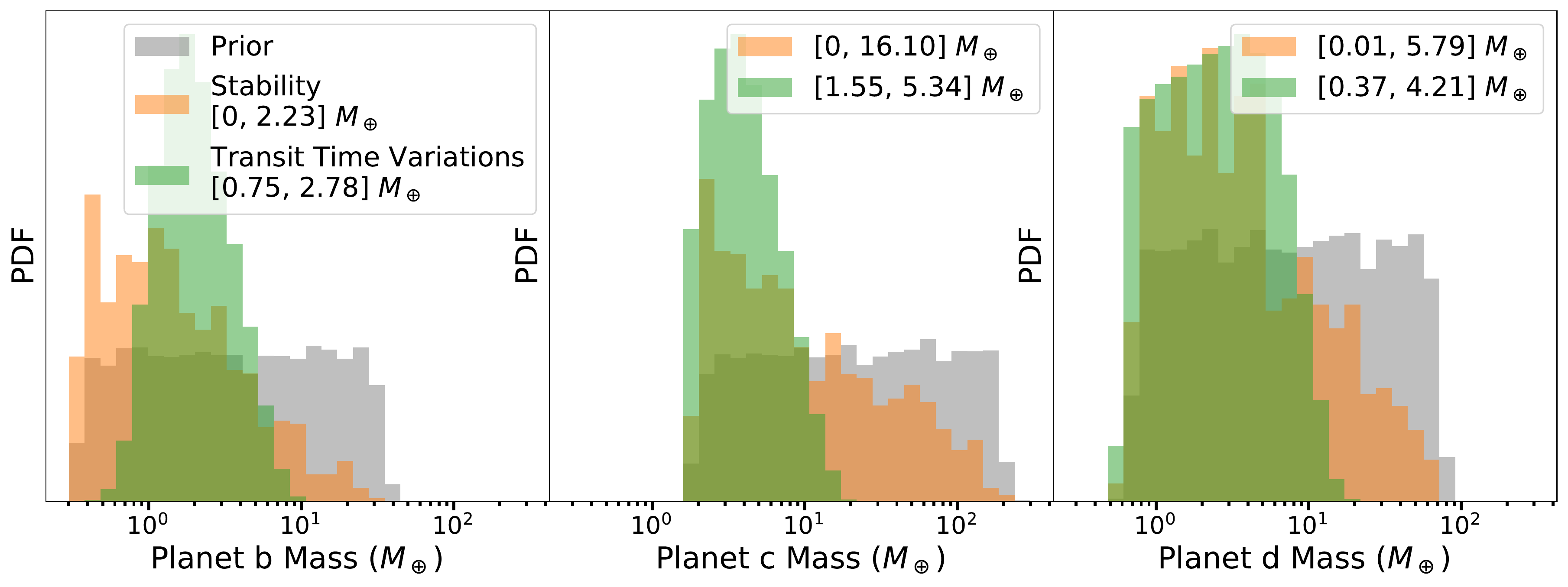}}
    \caption{Comparison of constraints on the masses of the three planets in the Kepler-23 systems from transit time variations and stability.
    Priors are plotted in gray.
    Following \protect\cite{Hadden17}, we also list in the legend the 68.3\% highest posterior density intervals, i.e., the smallest parameter range containing 68\% of the distribution.
    \label{fig:mcomp}}
\end{figure*}

In Fig.\:\ref{fig:mcomp} we compare the constraints on the planet masses, with 68\% highest posterior density intervals listed in the legend (see figure caption). 
Transit durations hold no information on planet masses, and we can see that TTVs provide stronger constraints than stability.
Additionally we point out that while TTVs can provide lower bounds on planet masses (as for planet $b$ in Fig.\:\ref{fig:mcomp}), requiring stability typically only provides upper limits.

We see that in this case, stability provides comparable upper mass limits to TTVs for the innermost and outermost planets (see legend of Fig.\:\ref{fig:mcomp}), but not for the middle planet.
We can understand this qualitatively.

At comparable planet masses, there will always be a preference toward lower-mass planets, which allow for stability over a broader range of orbital eccentricities.
However, in the limit where a single planet dominates the mass and the other bodies can be treated as test particles, the problem reduces to sets of analytically understood two-planet stability problems, which are much less restrictive.
As an illustration, a pair of planets on coplanar and initially circular orbits will never undergo close encounters as long as their separation is greater than $\approx 3.5$ Hill radii \citep{Gladman93}. 
This condition is satisfied for all pairs of planets in Kepler-23 even at the highest masses in the adopted prior. 
By contrast, equally separated 3+ planet systems (also on coplanar, and initially circular orbits) can undergo short-term instabilities out to separations that are roughly 3 times as wide \citep[e.g.,][]{Chambers96, Quillen11, Obertas17, Petit20}. Given that the Hill radius scales as $m^{1/3}$, this corresponds to mass constraints that differ by $\sim 30$ between two and three-planet cases.
In the Kepler-23 system, the middle planet has the largest radius, causing a preference for masses $>100 M\oplus$, which are excluded by the density priors on the other planets.
Stability constraints are therefore strongest in systems with comparable mass planets, which appears to be a common outcome in observed systems from investigations of correlations between planetary radii and masses in multiplanet systems \citep{Millholland17, Weiss18, Gilbert20, He20}.

Taking this multi-transiting system discovery, we can also ask what radial velocity precision would be required to reach the upper-mass limits imposed by stability. 
The 68\% upper mass limits of 2.2, 16.1, and 5.8 $M_\oplus$ correspond to radial velocity semi-amplitudes of 0.8, 4.8 and 1.5 m/s for planets $b$, $c$ and $d$, respectively.

\begin{figure*}
    \centering
    \resizebox{0.99\textwidth}{!}{\includegraphics{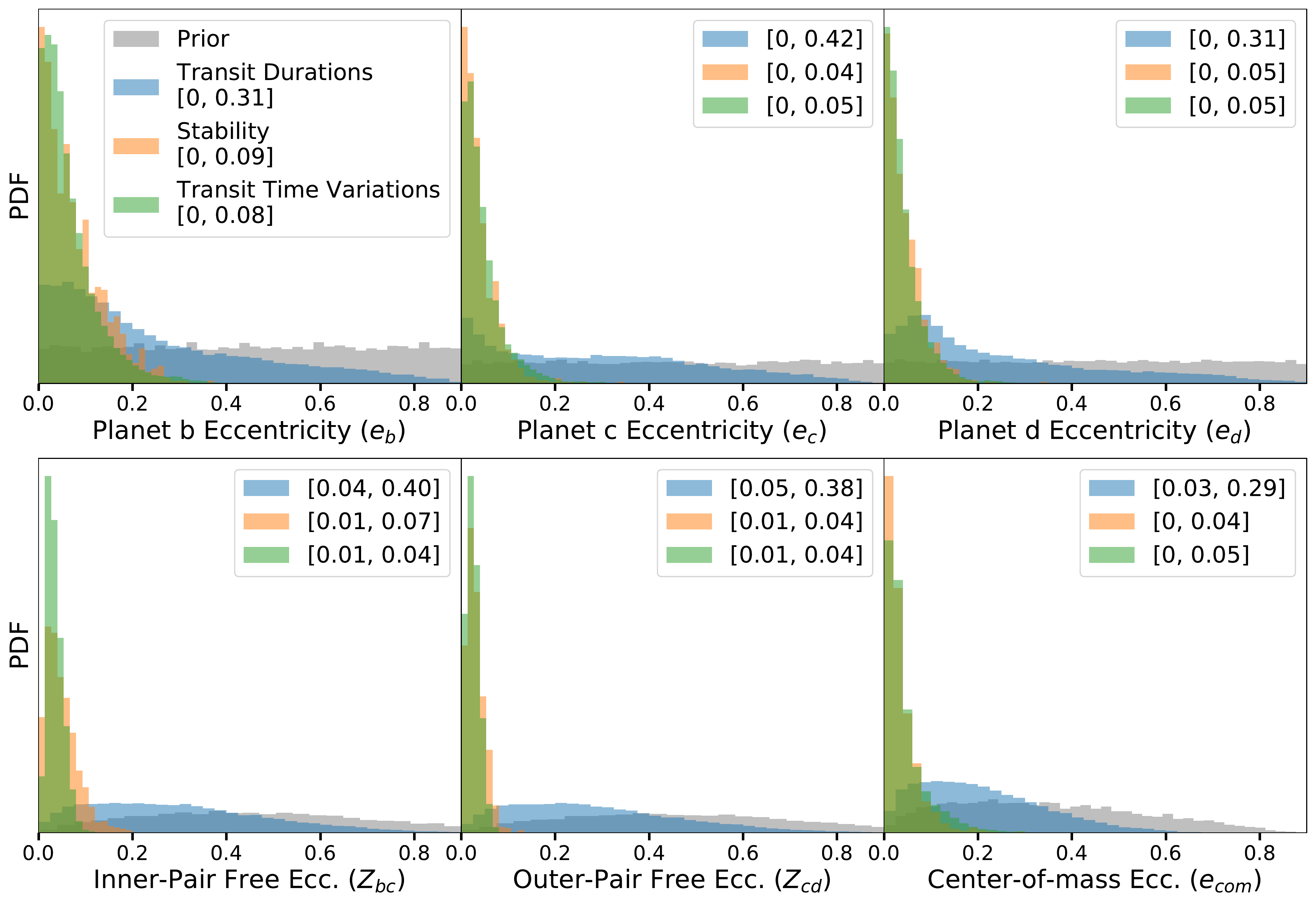}}
    \caption{ Top row: Comparison of constraints on the orbital eccentricities of the three planets in the Kepler-23 systems from transit durations, TTVs, and stability. Priors are plotted in gray. Following \protect\cite{vanEylen15} and \protect\cite{Hadden17}, we also list in the legend the 68.3\% highest posterior density intervals, i.e., the smallest parameter range containing 68\% of the distribution. Bottom row: Constraints on the particular eccentricity combinations $\mathcal{Z}$ (Eq.\:\ref{Z}) that drive the resonant dynamics between the inner (left panel) and outer (middle panel) planet pair. 
    The right panel plots a final combination of eccentricities that generalizes a conserved quantity in a two-planet, single MMR model (Eq.\:\ref{ecom}).
    \label{fig:ecomp}}
\end{figure*}

\subsection{Individual Eccentricities} \label{sec:methods}

We now compare the constraints on the orbital eccentricities for the Kepler-23 planets from transit durations \citep{vanEylen15}, TTVs \citep{Hadden17} and stability in the top row of Fig.\:\ref{fig:ecomp}, again listing the 68\% highest posterior density intervals. 
In this case, stability constraints approach those from TTVs, and are much stronger than those from transit durations.

Orbital eccentricities introduce radial velocity signals that are smaller than the observed semi-amplitude by a factor of $\sim e$ \citep[see, e.g.,][]{Shen08, Zakamska11}.
The 68\% eccentricity upper limits from stability of $0.05-0.1$ (Fig.\:\ref{fig:ecomp}) therefore correspond to radial velocity signals of $\lesssim 20$ cm/s for the masses found in Sec.\:\ref{sec:mass}.

\subsection{Free Eccentricities Between Adjacent Planet Pairs} \label{freeec}
One can show that for a single pair of planets near a j:j-k MMR\footnote{Each j:j-k resonance actually includes k+1 sub-resonances, but these can be combined into a single resonance via canonical transformation. This was originally shown (exactly) for first-order (k=1) resonances \citep{Sessin84, Wisdom86, Henrard86}, but \cite{Hadden19} shows that this also approximately holds for higher order resonances.}, there is a particular combination of the eccentricity vectors ${\bf \vec{e}}_i$ (with magnitude given by $e_i$ and direction given by the longitude of pericenter, $\varpi_i$) that is approximately conserved, while a second combination $\mathcal{Z}$ drives the resonant dynamics,
\begin{equation} \label{Z}
\mathcal{Z}_{i,i+1} \approx \Bigg| \frac{{\bf \vec{e}}_{i+1} - {\bf \vec{e}}_{i+1}}{\sqrt{2}}\Bigg|,
\end{equation}
where the exact expression carries additional coefficients for the eccentricity vectors that depend on the $j$ and $k$ indices of the resonance.
Except for the 2:1 MMR, these coefficients are within $\approx 10$\% of unity \citep[e.g.,][]{Deck13}.
The combination $\mathcal{Z}$ can further be decomposed into a component that is forced by the MMR, and a free component set by initial conditions. 
However, because most observed TTV systems are typically far from exact resonance ($\approx 1-5\%$ wide of the resonant period ratio), the forced component is typically negligible compared to the free component \citep[e.g.,][]{Hadden17}.
For the remainder of the paper we will therefore refer to values of $\mathcal{Z}$ as free eccentricities.

When sinusoidal variations in transit times measure eccentricities, they dominantly constrain these free eccentricities \citep[e.g.,][]{Lithwick12}.
TTVs thus provide narrower bounds on $\mathcal{Z}$ than on the individual eccentricities\footnote{In principle this is not a one-to-one comparison. If one orbit were eccentric and the other circular, $\mathcal{Z}$ would be a factor of $\sqrt{2}$ smaller than the non-zero eccentricity. However, for randomly oriented eccentricity vectors, like we assume in our priors, if both planets had orbital eccentricity $e$, then on average $\mathcal{Z} = e$.}.

At the same time, MMRs are also understood to drive fast instabilities in compact multiplanet systems \citep{Wisdom80, Quillen11, Deck13, Petit17, Hadden18, Petit20}.
Indeed, \cite{Tamayo20} showed that their machine learning model SPOCK, trained only in and near these discrete MMRs, generalizes to uniformly spaced systems.
Similar to TTVs, one would therefore expect stability to better constrain free eccentricities between pairs of adjacent planets than their respective individual eccentricities.
This has been demonstrated analytically for two-planet systems.
For higher multiplicity systems it is difficult to make precise statements without an analytic understanding yet in hand.
However, one would qualitatively expect that the closer a pair of planets are to a strong MMR, the more stability should specifically constrain the combination $\mathcal{Z}$, rather than the individual eccentricities.

Transforming from $N$ individual eccentricities to $N-1$ free eccentricities between adjacent pairs of planets leaves one additional degree of freedom. 
In the near-resonant two-planet case, one can show that there exists a conserved quantity, which (again to within near-unity coefficients) is equivalent to a center-of-mass eccentricity \citep[e.g.,][]{Hadden19}.
By analogy to this result, we define
\begin{equation} \label{ecom}
e_{\text{com}} = \frac{\Bigg| \sum_{i=1}^{N} m_i {\bf \vec{e}}_i \Bigg|}{\sum_{i=1}^N m_i},
\end{equation}
where the sum runs over all the planets. While this quantity is not strictly conserved, it reduces to an approximately conserved quantity in both the limit where a pair of near-resonant planets dominate the mass, and in the limit of a single dominant planet.

In the bottom row of Fig.\:\ref{fig:ecomp}, we show constraints on these transformed quantities.
We see that stability, and especially TTV, constraints on free eccentricities are tighter. 
In particular, TTVs are able to exclude zero from the 68\% interval.
We note that while the stability derived distributions now also peak at non-zero values, this is an artifact imposed by the adopted prior. 
Choosing eccentricity vectors randomly and independently results in few samples with comparable and aligned eccentricity vectors that would yield small values of $\mathcal{Z}$ (prior distributions in gray in Fig.\:\ref{fig:ecomp}).
This might approximate the random outcomes of giant impacts or planet-planet scattering, but would be a poor prior for, e.g., smooth migration in a disk with eccentricity-damping, which would act to damp free eccentricities to zero.
In this case a uniform prior in the free eccentricities would be more appropriate.

\begin{figure*}
    \centering
    \resizebox{0.99\textwidth}{!}{\includegraphics{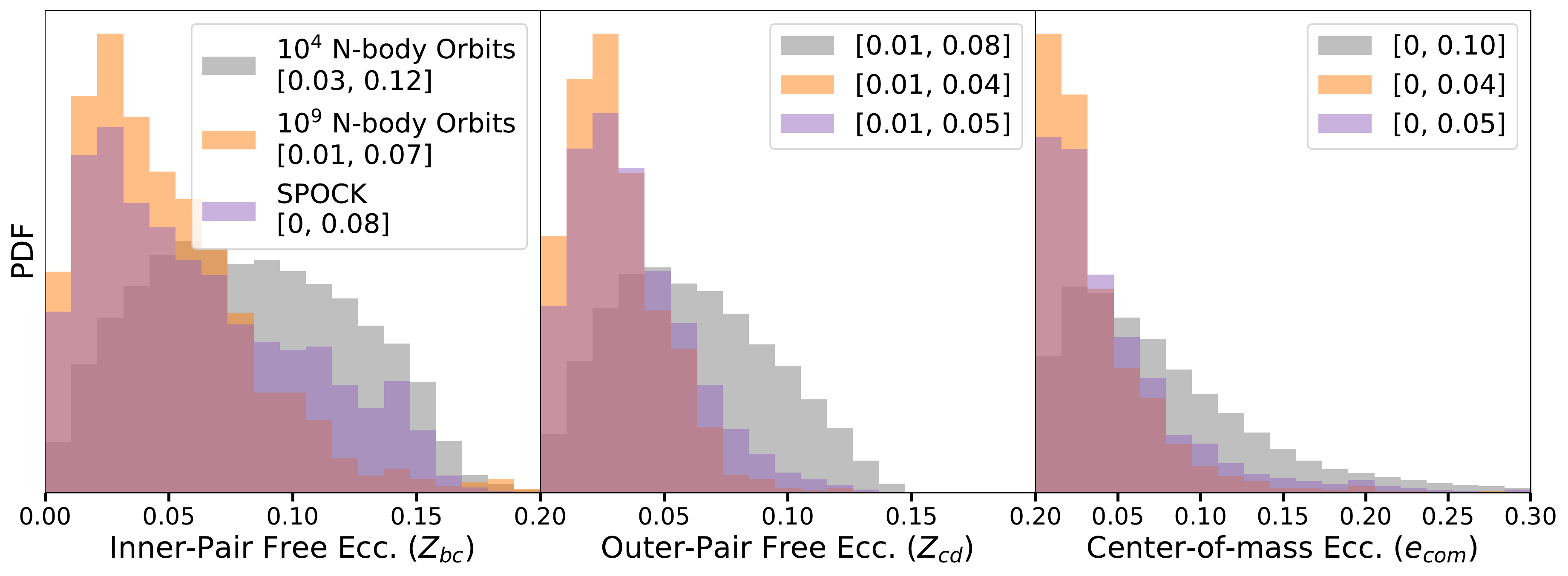}}
    \caption{ Comparison of the eccentricity distributions of configurations that remain stable over $10^9$ orbits in N-body integrations (orange) to faster estimates with SPOCK (purple). Gray histograms show configurations that remain stable over $10^4$ orbits. SPOCK \protect\citep{Tamayo20} effectively removes $96\%$ of the unstable configurations in the gray histogram to yield the purple histogram (see end of Sec.\:\ref{sec:spocknbodycomp}).
    \label{fig:spockcomp}}
\end{figure*}

Finally, we point out that while TTV modeling can rule out $\mathcal{Z} = 0$\footnote{Both the amplitude and phase of the TTV signal depend on $\mathcal{Z}$ \citep[e.g.,][]{Lithwick12}.}, stability will typically only yield upper limits, since things are generally more stable at lower eccentricities.
The exception would be near resonances that force non-zero equilibrium eccentricities.
In such cases, circular orbits far from the equilibrium could be strongly unstable and ruled out.
For example, \cite{Obertas17} found reductions in instability times by several orders of magnitude near MMRs in their numerical integrations of initially circular systems.

\section{Fast Stability Constrained Characterization} \label{sec:nbodycomp}

The N-body characterization above required approximately 8500 CPU hours in order to identify 833 stable configurations.
Given that typical MCMC analyses typically aim to have a factor of $\sim 100-1000$ more posterior samples, this renders N-body stability constrained characterization prohibitive for most applications.
We now consider two faster approaches: running shorter $10^4$-orbit N-body integrations to rule out only the fastest instabilities, and using SPOCK, the Stability of Planetary Orbital Configurations Klassifier \citep{Tamayo20}.
SPOCK is a machine learning model that also runs a direct $10^4$-orbit N-body integrations, but is trained to make stability predictions over $10^9$ orbits from this short time series \citep{Tamayo20}.

As justified below (Sec.\:\ref{sec:crossing}), we begin by rejecting orbit-crossing configurations, the vast majority of which should be unstable over long timescales.
In Fig.\:\ref{fig:spockcomp}, we compare the results of these faster approaches to the $10^9$-orbit N-body distributions labeled `Stability' in Figs\:\ref{fig:mcomp} and \ref{fig:ecomp}.
For the SPOCK predictions, we plot histograms of configurations weighted by their estimated probabilities of stability from the model following Eq.\:\ref{eq:sampapprox} (see Sec.\:\ref{sec:probabilistic}).

Of the 9414 configurations that survive $10^4$ orbits in gray, over 90\% go unstable when integrated to $10^9$ orbits.
We see in Fig.\:\ref{fig:spockcomp} that this leads to eccentricity constraints that are too wide by roughly a factor of two. 
SPOCK, by contrast, is able to match long-term N-body constraints to within approximately 20\% at comparable computational cost to the much shorter $10^4$-orbit integrations.
This illustrates SPOCK's potential for the fast characterization of compact multiplanet architectures.

We now further analyze SPOCK's performance to understand the source of deviations with the N-body distributions, and to make recommendations for its application to new systems.

\subsection{Probabilistic vs. Binary Classification} \label{sec:probabilistic}

For every input orbital configuration, SPOCK returns an estimated probability of stability.
For simplicity, \cite{Tamayo20} considered binary classification into stable and unstable systems, which one does by choosing a threshold probability to separate the two classes (Eq.\:\ref{binaryclass}).

We advocate instead for approximating $p(q\,|\,\theta)$ using SPOCK's continuous stability probability estimates directly.
While one can try to find a threshold that balances the rejection of most unstable systems without throwing out a significant fraction of the stable configurations one wishes to characterize, the right threshold will vary by system, and binary classification always throws out information.
By instead weighting all configurations by their stability probabilities estimated by SPOCK (Eq.\:\ref{eq:sampapprox}), we allow for more confident classifications from the model to be counted more heavily than ones close to a hand-tuned threshold.
This probabilistic approach also helps avoid potential pitfalls when using SPOCK on sets of configurations distributed very differently from its set of training examples.
We illustrate this through an instructive example.

\cite{Tamayo20} present a case study of the three-planet Kepler-307 system (third planet is a candidate), where they claim that SPOCK fails (their Fig.\:7).
In particular, they drew 1500 samples from the posterior of a TTV analysis of the system and generated SPOCK probabilities for each configuration.
In their earlier analysis, \cite{Tamayo20} held back 20\% of their training dataset.
Analyzing SPOCK's performance on this 20\% holdout set, they found that labeling systems with stability probabilities estimated by SPOCK $> 0.34$ lead to a false positive rate (FPR) of 10\%, i.e., only 10\% of unstable systems were misclassified as stable.

However, when they applied this same threshold to Kepler-307, this lead to an FPR of 87\%.
\cite{Tamayo20} argued that because the TTV analysis had already strongly constrained the resonant dynamics in the system, this caused the 1500 posterior samples to strongly cluster in SPOCK's feature space, making it difficult for the model to separate stable from unstable systems.
While this is certainly a factor, we now instead argue that this apparent failure is principally a result of their binary classification, and that a probabilistic approach resolves most of the disagreement.

\begin{figure*}
    \centering
    \resizebox{0.99\textwidth}{!}{\includegraphics{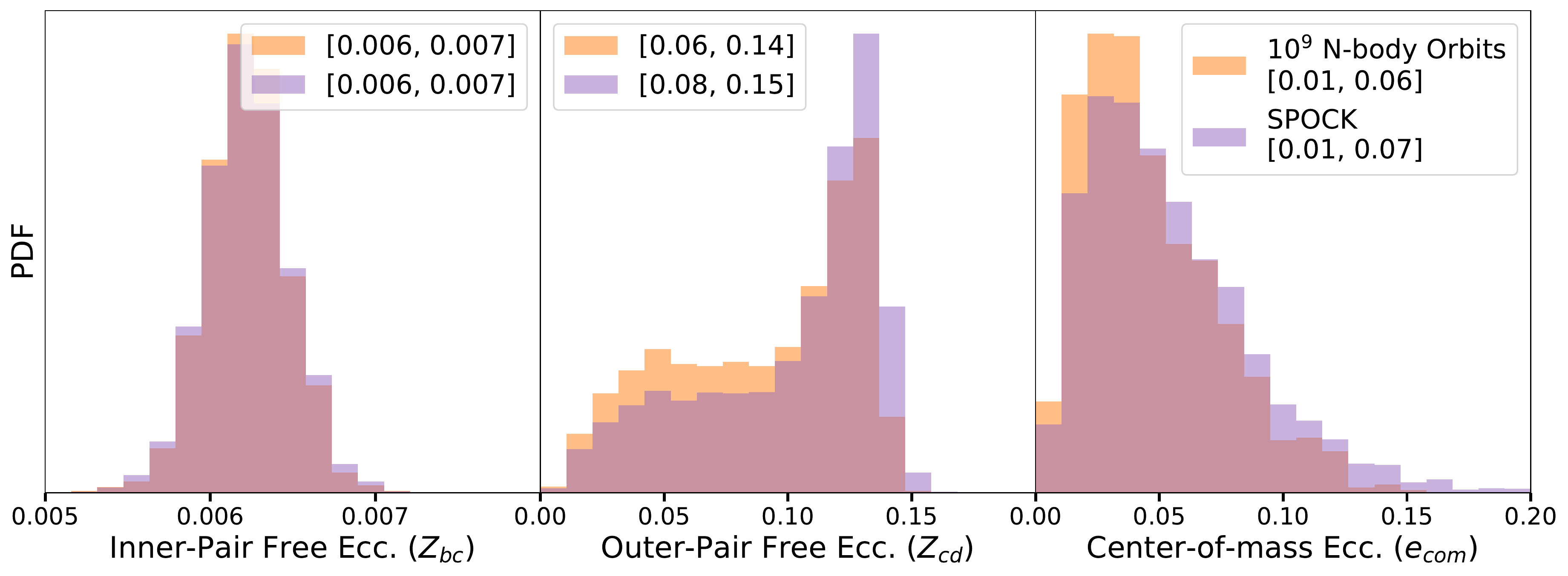}}
    \caption{ Separate test of SPOCK vs N-body on a different system, Kepler-307, presented by \protect\cite{Tamayo20} as a case where SPOCK fails. By continuously weighting configurations by their probabilities of stability as estimated by SPOCK (purple), we obtain significantly better agreement with stability constrained eccentricity posteriors performed with N-body integrations (orange) than by performing binary classification into stable and unstable configurations using a threshold optimized on SPOCK's training dataset as was done by \protect\cite{Tamayo20} (cf. their Fig.\:7).
    \label{fig:kepler307}}
\end{figure*}

For the sake of argument, let us assume that the probabilities of stability estimated by SPOCK are true probabilities.
In SPOCK's training set, there are many unstable configurations, which SPOCK assigns low stability probabilities.
The fact that SPOCK's FPR on this distribution of configurations is 10\% for a stability threshold of 0.34 means that 90\% of the unstable systems in its {\it training set} are assigned stability probabilities < 0.34.

But now imagine giving SPOCK a different set of much more stable orbital configurations, where the SPOCK probabilities are all $> 0.4$.
We should still expect many of these configurations to be unstable (e.g., only half of those with probability 0.5 should be stable).
But the fact that none of the systems in {\it this dataset} have stability probability below the 0.34 threshold, chosen using the {\it training dataset}, means that {\it all} of the unstable systems are misclassified as stable, i.e., the FPR is 100\%.

This is essentially what happened with Kepler-307.
The prior TTV analysis removed the vast majority of clearly unstable configurations, leaving over 95\% of samples with stability probabilities > 0.34 as estimated by SPOCK.
The fact that so few configurations fall below the threshold again means that the vast majority of unstable configurations are labeled as stable, yielding a misleadingly high FPR of 87\%.
In summary, a given probability threshold for binary classification is only sensible when the data distribution being tested is similar to the training data distribution.

The alternative is to weight configurations by their respective, continuous stability probability $p(q\,|\,\theta)$ as estimated by SPOCK (Eq.\:\ref{eq:sampapprox}).
We define the number of effective stable samples returned by SPOCK as the sum of SPOCK probabilities across the 1500 input systems.
This yields 1121 effective stable samples from SPOCK, as compared to 967 stable samples as determined through N-body integrations.
We therefore see that the probabilities estimated by SPOCK are overall too high, though the discrepancy is at the 10\% level.

In Fig.\:\ref{fig:kepler307}, we show the equivalent plot to Fig.\:\ref{fig:spockcomp} for Kepler-307, which also shows agreement with N-body to $\approx 20\%$. 
The most noticeable difference is that the N-body integrations can reject a larger fraction of systems with large free eccentricities between planets $c$ and $d$ than SPOCK.
As argued by \cite{Tamayo20}, we suspect that this is due to longer-term secular dynamics \citep[e.g.,][]{Hadden19} not effectively captured by SPOCK's features, which focus on the short-term resonant dynamics.
However, the agreement with N-body is much closer than suggested by the 87\% FPR reported by \cite{Tamayo20} using a binary classification threshold.
We therefore recommend weighting configurations by their stability probabilities as estimated by SPOCK.

Using continuous probabilities additionally makes the result differentiable with respect to orbital parameters, which allows for more efficient sampling in high-dimensional spaces through Hamiltonian Monte Carlo techniques \citep{exoplanet}.
SPOCK uses non-differentiable gradient-boosted decision trees, but we are working on differentiable models using neural networks (Cranmer et al., {\it in prep}).

\subsection{The Needle-In-A-Haystack Challenge}

The previous section considered an application of SPOCK where TTV constraints had already ruled out the most unstable configurations, leaving behind a mostly stable distribution of samples.
More interesting are cases where stability can rule out the majority of configurations to significantly constrain orbital parameters and masses, like in Kepler-23. 
It is instructive to first consider why such a task might be challenging for any stability estimator, including SPOCK.

In our analysis of Kepler-23, we sampled roughly 2 million orbital configurations, of which only 837, or $\sim 4$ in 10,000, were stable.
For the sake of argument, imagine SPOCK could correctly identify all 837 stable systems, which would exactly recover the orange N-body histograms in Fig\:\ref{fig:spockcomp}.
But now consider that, like any imperfect classifier, SPOCK suffers from some proportion of false positives, or unstable configurations it mislabels as stable.
Even for very low false positive rates (FPRs), say 1\%, the preponderance of unstable systems ($\sim 2 \times 10^6$) would lead one to include $\sim 20,000$ false positives in the predicted distribution of stable systems.
This would swamp the 837 stable systems, yielding unreliable results.
This textbook problem, often framed in terms of an imperfect medical test for a rare disease, arises across ``needle-in-a-haystack'' applications with strong class imbalances.
We also note that the most tightly packed systems, which provide the strongest stability constraints, by definition have the smallest fraction of possible stable configurations.
The most interesting systems for stability constrained characterization are therefore also the most challenging to accurately characterize.

\subsection{Rejection of Orbit-Crossing Configurations} \label{sec:crossing}

We found that one helpful way to alleviate this problem is to simply throw out configurations where any of the orbits cross.
Whether adjacent orbits cross is a non-linear problem involving the orbital orientations, but for closely spaced planets, the condition for the orbits of planet $i$ and $i+1$ to cross is simply \citep[e.g.][]{Hadden19},
\begin{equation} \label{ecross}
| {\bf \vec{e}_i} - {\bf \vec{e}_{i+1}} | > \frac{a_{i+1} - a_i}{a_{i+1}},
\end{equation}
where the $a$ denote semimajor axes, and $\vec{e} \equiv (e \cos{\varpi}, e \sin{\varpi})$ is a vector pointing in the direction of the longitude of pericenter $\varpi$, with magnitude given by the orbital eccentricity.

This is a reasonable cut given that planets on crossing orbits should eventually ``find" one another and scatter or collide; however, it is not a rigorous step. 
In particular, orbit-crossing configurations can be long-lived {\it in MMRs} (like those of Neptune and Pluto). 
One would therefore not want to take this step if there are reasons to suspect such a configuration, e.g., from short transit durations.
We note that in such cases, sampling eccentricities and pericenter orientations independently as done above will be very inefficient in sampling the resonant island \citep[see][for more effective sampling methods]{Tamayo20}.

In the typical case, however, where one can reasonably reject crossing configurations, this can significantly improve the accuracy of stability-constrained posteriors with SPOCK for two reasons.
First, the training dataset for SPOCK was generated drawing orbital eccentricities log-uniformly up to orbit-crossing values.
While SPOCK does generally assign low stability probabilities to crossing configurations, the lack of examples near and beyond orbit-crossing can lead to poor extrapolation.

More importantly, rejecting crossing orbits significantly cuts down the number of unstable configurations, alleviating the class imbalance challenge discussed above.
In this case, rejecting crossing configurations reduces the $\sim2$ million samples drawn from our prior by a factor of approximately 100. 
Having run all the N-body integrations, we can validate this rejection step.
We find that only 12 of these crossing configurations, or less than 1 in $10^5$, was stable over $10^9$ orbits.\footnote{In fact, manual examination of the 12 rejections that were long term stable reveal that while all were close, none were actually orbit-crossing---due to the leading order approximation in Eq.\:\ref{ecross}.
Given the negligible $1\%$ error on the estimated distributions, we do not pursue this correction.}

A more relevant comparison of these 12 outliers is to the total number of 837 stable configurations found.
However, the mis-rejection of these $1\%$ of stable configurations is more than compensated by the removal of $\approx 99\%$ of unstable systems, which we find would otherwise make up $\approx 20\%$ of the final estimated distribution (effective stable samples) in the form of false positives.

\subsection{Incorporating a Perfect Predictor} \label{sec:perfect}

SPOCK makes its stability predictions from summary features measured over a short N-body integration of $10^4$ orbits.
This allows SPOCK to catch systems that destabilize quickly, and rigorously assign them zero probability of stability over $10^9$ orbits\footnote{In reality the orbital evolution is chaotic, leading to a range of equally valid instability times, but the probability of one realization going unstable in $10^4$ orbits and another lasting $10^9$ orbits is negligible \citep{Hussain19}.}.
The fact that these highly unstable systems are classified perfectly, and do not contribute any residual probability to the estimated stable distribution as false positives further alleviates the class imbalance problem discussed above.

In total, rejecting crossing configurations removed $99\%$ of our unstable configurations, and the short $10^4$ orbit integrations from SPOCK further removed approximately 60\% as short-lived.
This left 9414 non-crossing configurations, surviving at least $10^4$ orbits, of which 825 $\approx 10\%$ are stable.
While the remaining class imbalance ratio of 10:1 is still a challenge, the improvement from the original ratio of 2500:1 is significant.

We note that while in principle the short integrations can also catch most of the unstable orbit-crossing configurations, the overwhelming number of crossing configurations causes the small fraction of surviving false positives to significantly skew the estimated distributions.
We therefore found that separately rejecting crossing configurations significantly improved predictions in this case with wide priors.

\subsection{Comparison between N-body and SPOCK} \label{sec:spocknbodycomp}

We now analyze the $\approx 20\%$ discrepancy between SPOCK and N-body shown in Fig.\:\ref{fig:spockcomp}.
In particular, are errors dominated by residual false positives, or are SPOCK probabilities significantly distorting the distribution of stable systems?
The former problem is significantly preferable to the latter.
After all, this is the baseline situation before applying any stability constraints, where {\it all} unstable configurations are included as false positives.
Thus, any approximate stability criterion is helpful (and conservative) as long as it removes unstable systems while preserving the distribution of stable configurations.

\begin{figure*}
    \centering
    \resizebox{0.99\textwidth}{!}{\includegraphics{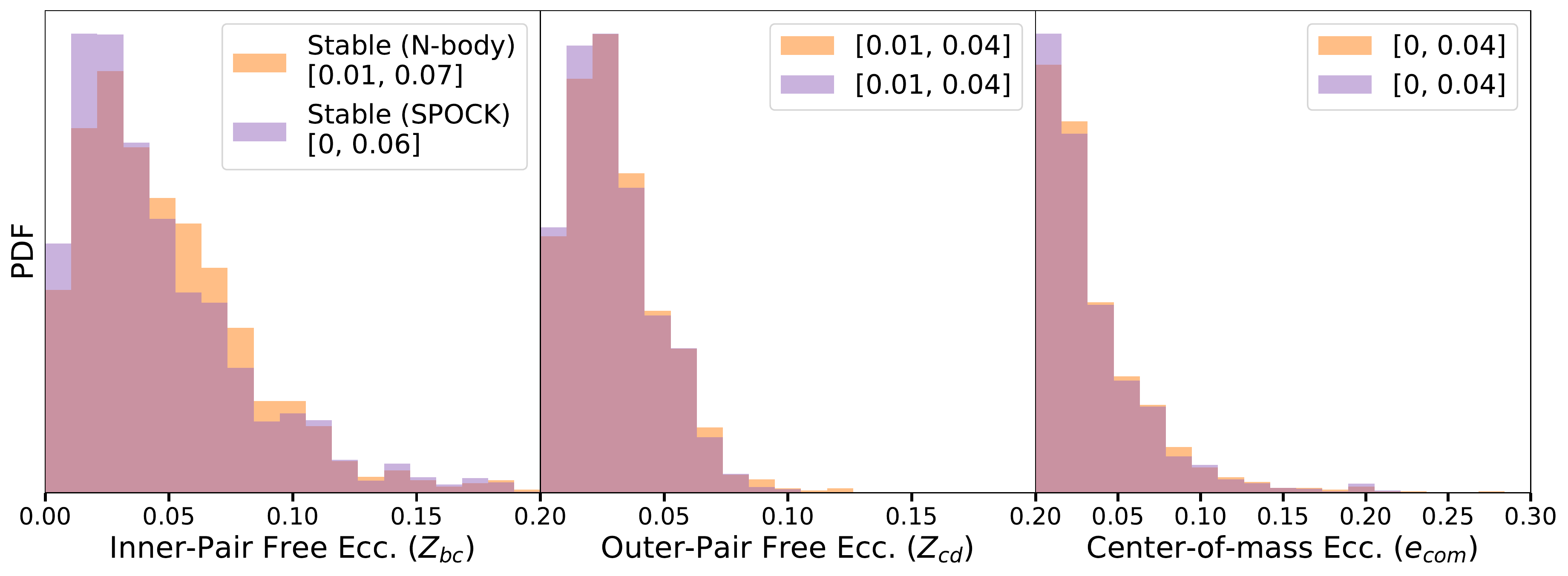}}
    \caption{ Comparison of the eccentricity distributions of only the configurations that remain stable over $10^9$ orbits in N-body integrations (orange) to the same histograms re-weighted by their estimated probabilities of stability estimated by SPOCK (purple). This shows that SPOCK is not significantly skewing the distribution of stable systems.
    \label{fig:spockstable}}
\end{figure*}

In Fig.\:\ref{fig:spockstable}, we plot histograms of configurations weighted by their SPOCK probabilities as in Fig.\:\ref{fig:spockcomp}, only taking the {\it stable} configurations as determined by N-body.
This demonstrates that SPOCK is not significantly distorting the distribution of stable systems we are trying to characterize.

We note that SPOCK does makes significant absolute errors in assigning these stable configurations stability probabilities below unity.
Starting from 833 stable configurations (orange), SPOCK returns 402 effective samples (purple).
However, it is not the absolute probabilities that matter, but rather the probabilities assigned to different samples relative to one another, and whether they distort the overall distribution.
We see in Fig.\:\ref{fig:spockstable} that the agreement is excellent, except for deviations in the free eccentricities between the inner pair of planets, where SPOCK skews the distribution by $\approx 15\%$ toward lower values. 

The final question is then how much SPOCK suppresses unstable systems.
As argued above, by assigning zero probability to systems that do not survive its short $10^4$ orbit integrations (and by rejecting orbit-crossing configurations), SPOCK can perfectly reject the vast majority of unstable configurations. 
The remainder are the 9414 samples plotted in the gray histogram in Fig.\:\ref{fig:spockcomp}, of which unstable systems outnumber stable ones 10:1.

SPOCK reduces these 8589 unstable configurations to 340 effective samples.
This final suppression of unstable systems by a factor of 25, while preserving the distribution of stable ones, allows SPOCK to faithfully approximate the underlying distribution of stable configurations in Fig.\:\ref{fig:spockcomp}.
This is a significant achievement in a problem where unstable configurations initially outnumbered stable ones 2500:1. 

\subsection{SPOCK Performance}

Because the stability probabilities from SPOCK approximate true probabilities, we have used the terms interchangeably above for simplicity.
In detail, however, this is not the case.
In particular, SPOCK probabilities are fit to capture the model's uncertainties on the distribution of example configurations it was trained on.
Consider creating a large sample of orbital configurations distinct from (but generated in the same way, with different random seeds, as) SPOCK's training examples \citep{Tamayo20}.
If one collected all samples where SPOCK estimates probabilities near 60\%, to excellent approximation, 60\% of those configurations would be stable over $10^9$ orbits when run with N-body, since this was what the model was trained to achieve.

However, while SPOCK's training set tries to cover the whole parameter range for observed near-coplanar compact multi-planet systems, any single system spans a much narrower range in phase space, where SPOCK might do better or worse than average.

\begin{figure}
    \centering \resizebox{\columnwidth}{!}{\includegraphics{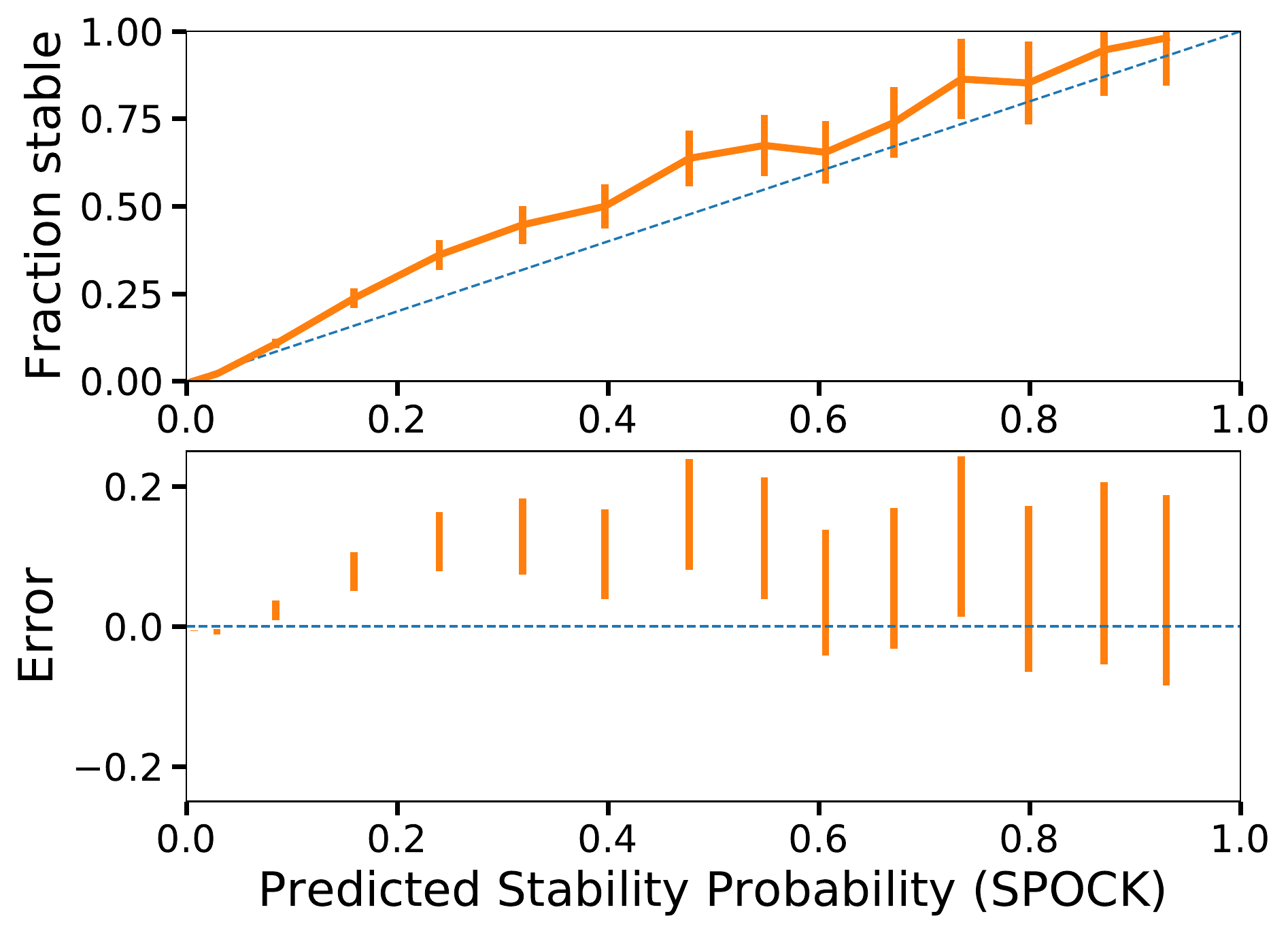}}
    \caption{ Predictions, binned by SPOCK's estimated probabilities of stability, vs. fraction of those configurations that were actually stable over $10^9$ orbits when run with N-body. A perfect model would follow the blue one-to-one line. Error bars denote Poisson counting errors, and bins were chosen so that each bin contributes the same total SPOCK probability to the final SPOCK distribution (purple histogram) in Fig.\:\ref{fig:spockcomp}.
    \label{fig:spockperformance}}
\end{figure}

In Fig.\:\ref{fig:spockperformance}, we evaluate SPOCK's performance on Kepler-23 by binning configurations by their probability of stability as estimated by SPOCK, and comparing it to the fraction of systems in the bin that were actually stable in N-body integrations (top panel). 
A perfect estimator would follow the 1:1 line in blue, and indeed, SPOCK achieves this when tested on holdout sets never seen during training of configurations drawn from its training dataset. 
For our sample of Kepler-23 configurations, we see that it deviates from the true probabilities by $\lesssim 20\%$ (bottom panel). 

Low probability configurations do not contribute much to the final distribution individually.
However, the preponderance of unstable systems means that the vast majority of configurations are assigned low stability probabilities.
Over 80\% of configurations have SPOCK probabilities < 0.1, as reflected by the small Poisson counting errors for the leftmost bins in Fig.\:\ref{fig:spockperformance}.

The bins in Fig.\:\ref{fig:spockperformance} were therefore chosen so that each bin contributes the same total probability to the final distribution.
This makes it a useful diagnostic for analyzing errors in the stability constrained distributions, since the errors in each bin then contribute equally to the purple histogram in Fig.\:\ref{fig:spockcomp}.

\subsection{Speed}

The $10^9$ orbit integrations of the 9414 non-crossing configurations that survive $>10^4$ orbits (Sec.\:\ref{sec:perfect}) required approximately 8500 CPU hours.
SPOCK predictions took $\approx 2$ CPU hours, a factor of 4000 improvement.
The precise speedup will depend on the particular configurations sampled.
For example, a sample of all stable configurations would be approximately $10^5$ times faster with SPOCK than N-body \citep{Tamayo20}.
By contrast, a sample of systems that all go unstable within $10^4$ orbits would take as long with SPOCK as with direct N-body given that SPOCK is performing a $10^4$ orbit N-body integration to generate its features.

Given that one does not know the instability time distribution for a given set of configurations ahead of time, it is difficult to estimate the time required through direct N-body integrations.
By contrast, SPOCK requires a flat evaluation time of $\approx 0.5$ CPU seconds per sample.

\subsection{Caveats and Future Work}

SPOCK was trained on resonant and near-resonant, compact three-planet configurations with masses $\lesssim 2$ Neptune masses, as typically observed in multiplanet exoplanet systems.
\cite{Tamayo20} present tests demonstrating accurate generalization to both non-resonant and higher multiplicity systems, which render SPOCK a useful tool for a wide range of exoplanet systems.

It would be valuable to test how far SPOCK's extrapolation can be pushed through comparison with N-body integrations of systems with, e.g., giant planets, and more inclined orbits (mutual inclinations $\gtrsim 10^\circ$).
One particularly interesting application is to resonant chains, where the resonant structure and chaotic separatrices can render wide swaths of parameter space unstable \citep[e.g.][]{Tamayo17}.
SPOCK's training dataset puts pairs of planets in and near mean motion resonances, but then initializes the third planet randomly. 
This means that very few training examples are in resonant chains, so SPOCK's performance on such cases is unclear.
At the same time, such resonant chains, e.g. TRAPPIST-1 \citep{Gillon17} and Kepler-223 \citep{Mills16} are both rare and particularly valuable objects of study, justifying large investments of CPU time with direct N-body.
An interesting direction would be to use suites of N-body integrations for such systems, and use transfer learning \citep[see, e.g.,][for a review]{Weiss16} to specialize SPOCK for stability classification for that particular case.

\subsection{When do stability constraints help?}

Without a full analytic understanding of stability in compact multiplanet systems, it is difficult to provide quantitative criteria for when stability constrained characterization will provide strong constraints.
Nevertheless, our partial understanding of the dynamics can provide some rough guidelines.

\cite{Quillen11} and \cite{Petit20} argue analytically that instabilities in compact, {\it initially circular}, 3+ planet systems are driven by 3-body resonances.
Most, if not all, observed multiplanet systems are at separations where 3-body resonances no longer overlap \citep{Quillen11, Petit20} and are stable against this mode of instability for plausible planet masses.
This is logical.
Systems unstable through this channel should have already eliminated themselves, so we should expect to find systems that are stable at zero eccentricity.

\cite{Tamayo20} suggest that short-lived eccentric configurations instead destabilize through the overlap of 2-body MMRs, given that their model trained on systems in and near such 2-body MMRs generalizes well to uniformly distributed compact configurations.
These resonant widths grow with increasing planet mass and orbital eccentricities \citep{Wisdom80, Deck13, Hadden18}, so one expects threshold combinations of masses and eccentricities beyond which a system will be short-lived, given a set of orbital periods (which are often the best-known parameters).

MMRs of a given order are more closely spaced (and thus easier to overlap) at closer separations (e.g., for first-order resonances, the period ratios of the 7:6 and 6:5 MMR are much closer than those of the 3:2 and 2:1).
Two-body MMRs always overlap as eccentricities reach orbit-crossing configurations \citep{Hadden18, Hadden19}, so one can expect stable configurations to extend to a fraction of orbit-crossing values, where the exact threshold will be correlated with the planet masses.

Perhaps most importantly, constraints will depend strongly on the relative spacing between adjacent planets. 
As argued in Sec.\:\ref{sec:mass}, numerical experiments suggest that mass constraints can vary by over an order of magnitude between two-planet cases and ones with 3+ equally spaced planets.
Thus, the strongest constraints will not necessarily occur in the system with the closest pair of adjacent planets, but in ones with comparably spaced additional bodies.
Interestingly, transiting multiplanet systems exhibit a preference toward similar period ratios between adjacent planets \citep{Millholland17, Weiss18}.

We have seen that in cases like Kepler-23 where both pairs of planets have period ratios at or less than 3:2, one can obtain strong constraints even for planet masses in the super-Earth regime.
The value of SPOCK is that one can quickly check stability constraints on estimated orbital parameters and masses, whereas N-body integrations are typically prohibitive.

\section{Conclusion} \label{conclusion}

Kepler-23 is a particularly informative compact three-planet system, where mass and eccentricity constraints from transit timing variations (TTVs), transit durations and stability can be directly compared.
Through this case study we have shown that in the most compact systems, stability can provide comparable constraints to TTVs, and much narrower upper limits than transit durations (Figs.\:\ref{fig:mcomp} and \ref{fig:ecomp}).
This is particularly relevant for multiplanet systems observed over short time baselines with only a few transits, where TTVs rarely provide strong constraints \citep{Hadden19}.

Such stability constrained characterization is typically prohibitive through direct N-body integrations, due to the long system ages and the large number of orbital configurations to be evaluated.
We compared these computationally expensive N-body constraints to much faster ones using the Stability of Planetary Orbital Configurations Klassifier (SPOCK) from \cite{Tamayo20}.

Stability constrained characterization with approximate stability estimators like SPOCK is challenging for compact systems where most candidate configurations are unstable (drawing from our prior resulted in a ratio of 2500 unstable systems for every stable one).
Given the preponderance of unstable configurations, in order for residual false positives not to dominate the final posteriors requires strong rejection of unstable systems without skewing the distribution of stable ones.
We show that SPOCK indeed approximately preserves the distribution of stable configurations (Fig.\:\ref{fig:spockstable}), and strongly suppresses unstable configurations to yield good agreement with N-body (Fig.\:\ref{fig:spockcomp}).

We provide several recommendations for stability constrained characterization with SPOCK, most notably first rejecting any crossing configurations (Sec.\:\ref{sec:crossing}), and weighting individual configurations by their respective probabilities of stability as estimated by SPOCK (Sec.\:\ref{sec:probabilistic}).

While contrasting constraints from transit timing variations, transit durations and stability is informative, we emphasize that they are not meant to be mutually exclusive, and are in fact complementary (Sec.\:\ref{sec:scc}).
For example, as the GAIA mission increases the sample of stars with well measured densities beyond the asteroseismic sample \citep{vanEylen15}, joint analyses of multiplanet systems with transit durations disfavoring low eccentricities and stability cutting off large values could provide improved constraints.

\section*{Acknowledgements}

We thank the anonymous reviewer for a careful report that improved the presentation in this manuscript.
We would also like to thank Vincent van Eylen and Sam Hadden for sharing their data on Kepler-23 for comparison. 
DT is indebted to the Hews Company for IT support that enabled the completion of this work remotely.
Support for DT was provided by NASA through the NASA Hubble Fellowship grant HST-HF2-51423.001-A awarded  by  the  Space  Telescope  Science  Institute,  which  is  operated  by  the  Association  of  Universities  for  Research  in  Astronomy,  Inc.,  for  NASA,  under  contract  NAS5-26555.
C.G. acknowledges support from the Penn State Eberly College of Science, Department of Astronomy \& Astrophysics, the Center for Exoplanets and Habitable Worlds, the Center for Astrostatistics, Pennsylvania Space Grant Consortium, and the Institute for Computational and Data Sciences (\url{http://ics.psu.edu/}) at The Pennsylvania State University, including the CyberLAMP cluster supported by NSF grant MRI-1626251, for providing advanced computing resources and services that have contributed to the research results reported in this paper.
The citations in this paper have made use of NASA's Astrophysics Data System Bibliographic Services.
This research has made use of the NASA Exoplanet Archive, which is operated by the California Institute of Technology, under contract with the National Aeronautics and Space Administration under the Exoplanet Exploration Program.
This research was made possible by the open-source projects \texttt{REBOUND} \citep{Rein12}, SPOCK \citep{Tamayo20},
\texttt{Jupyter} \citep{jupyter}, \texttt{iPython} \citep{ipython}, \citep{numpy},
and \texttt{matplotlib} \citep{matplotlib, matplotlib2}.

\section*{Data Availability}
To aid in future stability analyses, we provide an accompanying repository with Jupyter notebooks to reproduce the figures in this paper at \url{https://github.com/dtamayo/Stability-Priors}.




\bibliographystyle{mnras}
\bibliography{Bib}






\bsp	
\label{lastpage}
\end{document}